\documentclass[12pt]{article}

\def\beq{\begin{eqnarray}}
\def\eeq{\end{eqnarray}}

\def\mpl{M_{\rm Pl}}

\def\bo{{\nabla^2}}
\def\fu{{\cal F}}

\def\lsim{\mathrel{\rlap{\lower3pt\hbox{\hskip0pt$\sim$}}
    \raise1pt\hbox{$<$}}}         %less than or approx. symbol
\def\gsim{\mathrel{\rlap{\lower4pt\hbox{\hskip1pt$\sim$}}
    \raise1pt\hbox{$>$}}}         %greater than or approx. symbol

\begin{document}

\begin{flushright}
HUTP-02/A047  \\
Stanford--ITP-02-37\\
NYU-TH/02/09/10 \\
CERN --TH/2002-247\\
~~~\\
September 26, 2002 \\
\end{flushright}

\vskip 1cm
\begin{center}
{\Large \bf
Non-Local Modification of Gravity and the Cosmological Constant Problem
\vskip 0.2cm}
\vskip 1cm
{Nima Arkani-Hamed$^a$, Savas Dimopoulos$^b$, Gia Dvali$^c$,
Gregory Gabadadze$^d$}

\vskip 1cm
{\it $^a$Jefferson Laboratory of Physics, Harvard University,
Cambridge, MA 02138 \\
$^b$Physics Department, Stanford University, Stanford, CA  94305 \\
$^c$Department of Physics, New York University, New York, NY 10003 \\
$^d$Theory Division, CERN, CH-1211  Geneva 23, Switzerland}\\
\end{center}

\vspace{0.9cm}
\begin{center}
{\bf Abstract}
\end{center}
We propose a phenomenological approach to the cosmological constant
problem based on generally covariant  non-local and acausal
modifications of four-dimensional gravity at enormous distances.
The effective Newton constant becomes
very small at large length scales, so that sources with immense wavelengths
and periods --- such as the vacuum energy--- produce miniscule
curvature. Conventional astrophysics, cosmology and standard
inflationary scenaria are unaffected, as they
involve shorter length scales. A new possibility emerges that
inflation may ``self-terminate''
naturally  by its own action of stretching wavelengths
to enormous sizes.  In a simple limit our proposal
leads to a modification of Einstein's equation by a single additional
term proportional to the average space-time curvature of the Universe. It
may also have a qualitative connection with the dS/CFT conjecture.

\vspace{0.1in}

\newpage

\section{Introduction}

\vspace{0.5cm}

The Cosmological Constant Problem (CCP) is one of the most pressing
conceptual problems in physics. The energy-momentum tensor $T_{\mu \nu}$
is expected to contain a vacuum energy density
piece ${\cal E} g_{\mu \nu}$, and the
natural value for $\cal E$ coming from the Standard
Model sector should be
at least $\sim ({\rm TeV})^4$. However, according to the Einstein
equations
\begin{equation}
M_{\rm Pl}^2 \, G_{\mu \nu} \,= \,T_{\mu \nu}\,,
\end{equation}
such a vacuum energy would give rise to a drastically different cosmology
than what we observe. For instance, if ${\cal E} > 0$, the universe
quickly
becomes asymptotically de Sitter with a radius of curvature $\sim$ mm,
while the observed curvature radius of the
Universe is enormously larger, $\sim H_0^{-1} \sim 10^{28}$ cm.

The most familiar formulation of the CCP is ``Why is the
vacuum energy so small?". This reflects the most common approach to the
problem: to invoke some dynamics, analogous to the Peccei-Quinn
mechanism for the strong CP problem,
that is flexible enough to adjust and cancel any value of
the vacuum energy (see for a review Ref. \cite {Weinberg}).
In this formulation, the mystery is even further
deepened by recent cosmological observations \cite {cc}
suggesting that the Universe has recently entered an accelerated
phase with the curvature radius $\sim
H^{-1}_0$, which is usually ascribed to a tiny
${\cal E} \sim ({\rm mm})^{-4}$.
The question then becomes ``why is the vacuum energy density
so tiny, 60 orders of magnitude smaller than its natural value,
but not zero''?

However, a more precise formulation of the problem is: ``Why does the
vacuum energy gravitate so little?".
This suggests an approach where ${\cal E}$ keeps its natural value
$\sim ({\rm TeV})^4$, but gravity is modified so that this large vacuum
energy density does not give rise to large observable curvature
(for attempts see \cite{lindecc, tseytlin}).

More specifically, it is natural to try
and modify gravity in the {\it infrared} (IR) to address the CCP, given that
the CCP seems to be associated with very low energy scales
with respect to the Planck scale or even the weak scale.
However, until recently there were no explicit examples
of consistent theories where the behaviour of
gravity is classically modified at large distances.
This situation changed with the
advent of models where the Standard Model fields are localised to a brane in
infinite volume extra dimensions \cite{dgp}, where gravity on the brane
transitions from being four-dimensional to higher-dimensional at very large
distances.  These models motivated afresh
the possibility of addressing the cosmological constant problem by IR
modification of gravity \cite{gg}, where the vacuum energy (brane tension)
mostly curves the bulk, while ordinary gravity is trapped to the brane at
observable distances by the
presence of a large Einstein-Hilbert action localised on the brane.
A specific proposal along these lines
was made in Ref. \cite{dgs}, where it is argued that the
graviton propagator is modified in the
infrared in such a way that
large wavelength sources, such as the vacuum  energy, gravitate very weakly.
As a result, even huge vacuum energy does not curve space.
On the other hand, short wavelength sources, such as
planets, stars, galaxies and clusters gravitate
(almost) normally.

It is challenging to perform computations
in the framework of \cite{dgs} in order to explore its consequences in
a realistic setting. However the general idea of
addressing the CCP with non-local modification of gravity in the
infrared is so attractive that it is desirable to
try and explore the viability of this idea,
as well as the properties such non-local modifications should have, in a
concrete way.

This leads us to ask a more modest question:
What should an
effective, four-dimensional, long-distance, classical description
of physics incorporating
non-locality look like, in order to address the CCP in a realistic setting?

\section{Newton's Constant as a High-Pass Filter}

The fundamental physical idea we want to implement, inspired by
the example of \cite{dgs}, is to make the effective Newton constant
depend on the frequency and wavelength in such a way that for sources that
are uniform in both space and time, such as the vacuum energy,
the effective Newton constant is tiny, shutting off their
gravitational effects. Analogs in electromagnetism are
frequency-dependent dielectrics and high-pass filters.

We will not attempt to
derive this physics from a consistent quantum effective field theory.
Indeed, the modifications we will end up considering
are non-local and acausal, and are hard to imagine coming from conventional
field theories. Instead, we will simply modify
Einstein gravity at the level of classical
equations of
motion, implementing the
physical idea of ``the Newton constant
as a high-pass filter'', with the goal of
resolving the CCP without any
fine adjustments of the parameters in the equations of motion, while
retaining all the usual successes of general relativity. As we will see,
this can be accomplished with extremely simple non-local modifications of
the equations of motion. We will not attempt
to derive our equations of motion from a variational principle; this is
no great loss since any action principle itself would have to be non-local
and would not necessarily directly lead to a sensible quantum theory in any
case. Also the example of \cite{dgs} suggests that the {\it local}
action formulation may require going beyond the four
dimensional theory.

A first example of a modified Einstein equation
that incorporates the above-mentioned properties takes the form
\beq
\mpl^2\,\left (  1\,+\,\fu (L^2\,\bo ) \right )\,G_{\mu\nu}\,
=\,T_{\mu\nu}\,,
\label{1}
\eeq
where  $\fu(L^2\,\bo)$ is the ``filter function''
with the following properties:
\beq
\fu (\alpha ) \to 0\, ~~~{\rm for}~~~ \alpha \gg 1\,;
\label{f1} \\
\fu (\alpha )\, \gg \, 1 \,~~~{\rm for}~~~ \alpha \ll 1\,.
\label{f2}
\eeq
$L$ in (\ref {1}) is a distance scale at which gravity
is modified; it can be infinite, or very large but finite;
$\bo \equiv \nabla_\mu \nabla^\mu$
denotes the covariant d'Alambertian.
One can think of (\ref {1}) as the Einstein equation
with the effective Newton constant $(8\pi G_N^{\rm eff})^{-1}=
\mpl^2 (  1+\fu )$. It is immediately clear that at
least for the case where $T_{\mu \nu}$ is pure vacuum energy density
${\cal E} g_{\mu \nu}$, the maximally symmetric solution to
these equations of motion can have acceptably small
curvature if $\fu(0)$ is large enough. This is because in a maximally
symmetric space $G_{\mu \nu} = -g_{\mu \nu} {R}/{4}$,
where $R$  is the (space-time constant)
Ricci scalar. Since $g_{\mu \nu}$ is covariantly
conserved, $\nabla^2 g_{\mu \nu} = 0$, so
\begin{equation}
M_{\rm Pl}^2\,\left ( 1 + \fu(L^2 \nabla^2) \right )\, G_{\mu
\nu}\, = \, \left ( M_{\rm Pl}^2 + \bar{M}^2 \right ) \,
G_{\mu \nu}\,,
\end{equation}
where
\begin{equation}
{\bar M}^2\,=\,\fu(0)\,\mpl^2\,\gg \, \mpl^2\,,
\label{Mbar}
\end{equation}
and therefore
\begin{equation}
R = -\frac{4 {\cal E}}{M_{\rm Pl}^2 + \bar{M}^2}\,,
\end{equation}
which can be sufficiently small provided that $\bar{M}$ is sufficiently
large. Note that this does not require any fine adjustments for ${\cal E}$
or $\bar{M}$ so long as $\bar{M}$ is large enough.

It is also instructive to see how this effective suppression of the
cosmological constant is seen in a linearised approximation about flat
space; usually, the ${\cal E}$ generates a tadpole for the graviton.
The effect of the $\fu(L^2 \nabla^2)$
term is to modify the graviton propagator in momentum space to
(neglecting
indices) $(1 + \fu (k^2L^2))^{-1} 1/k^2$; as $\fu(0)$ is made large, this
shuts off the propagator at zero external momentum and removes the effect
of the tadpole completely in the $\fu(0) \to \infty$ limit.
This is a generalisation of the propagator of Ref. \cite{dgs}.
This is not surprising, since, as we said above,
this theory modifies gravity in far infrared.

It is also interesting that this kind of
modified propagator can arise without using branes or extra dimensions in
an important way, but instead by perturbatively
modifying the world-sheet action
in string theory \cite{eva}. Such modifications are used
to eliminate the effect of tadpoles generated in non-SUSY string
theories, leading to new perturbative stringy
backgrounds which are static and
non-supersymmetric \cite{eva}.

\section{$L \to \infty$}

\subsection{A Concrete Example}
After this motivation, we now examine the physical consequences of this
idea in more detail. To do so, it is convenient to consider the simplest
possible modification with the desired properties, which we can loosely
motivate by taking the $L\to \infty$ limit in Eq. (\ref {1}).
Roughly speaking, the  $\fu(L^2 \nabla^2) G_{\mu \nu}$ piece will then
extract the space-time ``zero mode'' of $G_{\mu \nu}$; a zero mode
$\psi_{\mu \nu}$ would satisfy $\nabla^2 \psi_{\mu \nu} = 0$, which
always
has a generic solution $\psi_{\mu \nu} = g_{\mu \nu}$ since $g_{\mu \nu}$
is covariantly constant. (For our heuristic purposes here we assume that
we are working with a Euclidean metric; for Minkowski space there are
additional solutions corresponding to the excitation on the light-cone).
So as $L$ goes to infinity, we can
replace $\fu (L^2 \nabla^2) G_{\mu \nu}$ by the ``zero mode''
part of $G_{\mu \nu}$, which is proportional to
$g_{\mu \nu}$. The constant of proportionality can be determined
by taking trace, and we arrive at the equation of motion

\beq
\mpl^2\,G_{\mu\nu}\,-\,{1\over 4}\,{\bar M}^2\,g_{\mu\nu}
{\bar R}\,=\,T_{\mu\nu}\,,
\label{rbar}
\eeq
where
\beq
{\bar R}\,\equiv \, { \int \,d^4x\,\sqrt{g}
\,R \over \int \,d^4x\,\sqrt{g}}\,
\label{rbardef}
\eeq
is the space-time averaged Ricci curvature
and ${\bar M}$ is defined in (\ref {Mbar}).

The above ``derivation'' was mainly intended to motivate this equation of
motion. It is interesting that Eq. (\ref {rbar})
is universal, independent of the filter function $\fu$.
We could have heuristically argued for
(\ref {rbar}) as follows:
in the limit  $L\to \infty $ the filter
excises only the infinite wavelength and
period fluctuations from the dynamics.
These involve  the space-time Fourier transform
at vanishing momentum and frequency, i.e., the space-time average
of dynamical quantities.
General covariance dictates that we only consider space-time averages
of scalars. Furthermore, the simplest dynamical scalar
in gravity is the curvature scalar. This suggests that we
modify Einstein's equations by a term proportional to the
space-time average of the curvature scalar.
But this is precisely what Eq. (\ref {rbar}) does.
The electromagnetic analogue of  Eq. (\ref {rbar}) is:
\beq
\nabla \cdot  {\vec E}\,+\,{\bar \epsilon}}\, {\overline {
\nabla  \cdot  {\vec E}} \,=\,4\pi \,\rho\,,
\label{dielectric}
\eeq
where $1+{\bar \epsilon}=1+\epsilon (\omega =0)$ is the dielectric
constant at zero frequency, which,
if large, suppresses the effects of homogeneous charge fluctuations.
In any case, we could have started by
postulating Eq. (\ref {rbar}), and we will shortly consider yet another
motivation
for considering this equation of motion, in the context of a
possible connection with the dS/CFT conjecture.

Note that the equation of motion (\ref {rbar})
is consistent in the sense that it manifestly satisfies the Bianchi
identities; this is true because for a
given space-time $\bar{R}$ is simply a number, and therefore (since the
metric is covariantly constant) the covariant
divergence of the l.h.s. of the equation vanishes.

Clearly the definition of $\bar{R}$ is a formal one. There can
be divergences in the integration both in the numerator and denominator,
in infinite space-times or in the presence of
short-distance curvature singularities. We can deal with the latter
by excising regions of Planckian curvature from the space-time in the
integrations. The former ambiguity can be dealt with
in a broad class of examples. Clearly $\bar{R} = 0$ in
any space-time with infinite volume but a finite integrated Ricci scalar. This
includes any asymptotically flat space-time sprinkled with stars or
black holes. $\bar{R}$ is also zero in a radiation, or matter-dominated,
forever expanding
Friedmann-Robertson-Walker (FRW)
universe, because $\int dt R(t)$ converges (away from the big-bang
singularity). In maximally symmetric dS or AdS spaces, $R$
is constant, the numerator in (\ref {rbardef}) is $R$ times the
denominator, so it makes
sense to define $\bar{R} = R$. Now consider a space-time that begins with a
big-bang and is asymptotically dS with de Sitter curvature $R_{\infty}$.
Both the numerator and denominator are completely dominated by the (infinite)
contributions from the asymptotic dS region in the future and therefore,
in these space-times, $\bar{R}$ is reasonably defined to equal $R_{\infty}$.
These examples suffice for our immediate purposes, but it may turn out that
more general prescriptions for defining $\bar{R}$ are needed in more
interesting space-times.

Our modified equation of motion coincides with Einstein's equation
for any system for which $\bar{R}=0$. As we have just discussed
this includes localised solutions such as a star, black hole,
and
matter- or radiation-dominated FRW cosmologies. Its main
new physical consequence is that the enormity of
$\bar M$ suppresses the value of $\bar R$,
in spaces with non-vanishing $\bar R$, such as de Sitter space.
This is accomplished without finely adjusting any of the parameters in the
equation of motion.
For simplicity, let us consider an energy-momentum tensor $T_{\mu \nu}$
that consists of a vacuum energy piece together with another
contribution
from radiation and non-relativistic matter
\beq
T_{\mu\nu}\,=\,g_{\mu\nu}\, {\cal E}\,+\,T^{\rm other}_{\mu\nu}\,.
\label{Tdec}
\eeq
Now, let us restrict our attention to space-times that begin with some
generic big-bang singularity but are asymptotically de Sitter in the
future. As we have argued, in such space-times, $\bar{R}$ is given by the
asymptotic dS curvature, $R_{\infty}$.
We can self-consistently compute $\bar{R}$ by looking at the
equation of motion in the deep future, where all sources of
energy-momentum other than the vacuum energy
have inflated away. Then we
conclude that
\begin{equation}
{\bar R}\, = R_{\infty} = \,\frac{-4 \,
{\cal E}}{M_{\rm Pl}^2 + {\bar M}^2}\,.
\label{cc0}
\end{equation}
This is a tiny curvature if we make ${\bar M}^2$ large
enough. For ${\cal E}$
near its smallest value
compatible with naturalness, $\sim$ (TeV)$^4$, we need
$\bar{M} \sim 10^{48} \, \, \mbox {GeV}$
in order to reproduce
the observed acceleration of the Universe today. If ${\cal E}$ has the
largest
size $\sim M_{{\rm Pl}}^4$, then we need $\bar{M} \sim 10^{80}$ GeV,
which is the mass of the Universe!

One may have thought that
a natural value for ${\bar M}$ would be close to
$M_{\rm Pl}$, but there is in
fact no reason to believe this: $M_{\rm Pl}$ sets a short-distance physics
scale, where gravity gets strongly coupled, while ${\bar M}$ clearly has
to do with
deep infrared, non-local physics. We already know that there is a
large hierarchy
between the weak scale and the Planck scale, so there may be a hierarchy
between $M_{\rm Pl}$ and $\bar{M}$; it is of course tempting to speculate
that in a fundamental theory these hierarchies are related. However, this
large value of $\bar{M}$ does not need to be finely adjusted to  any
particular value in any sense.

We can not address the full issue of the radiative stability of these
parameters since we do not  have
a full quantum theory of gravity; however, since we are modifying only the
gravitational part of the equation of motion we can
discuss radiative stability at least at the level of Standard
Model radiative corrections.  We assume  $T_{\mu \nu}$
on the r.h.s. of the equation of motion to be derived in the usual way
from the Standard Model
quantum effective action $\Gamma_{\rm SM}[g]$, evaluated in a gravitational
background $g_{\mu \nu}$:
\begin{equation}
T_{\mu \nu}\, = \,\frac{2}{\sqrt{g}}\, \frac{\delta \Gamma_{\rm SM}}
{\delta g^{\mu \nu}}\,.
\label{TSM}
\end{equation}
The various terms in
$\Gamma_{\rm SM}$ are assumed to have their natural sizes.
Since $\Gamma_{\rm SM}$ is
obtained from quantum field theory loops and is completely local, there is
no renormalization of ${\bar M}$ from Standard Model
loops. In our considerations so far we have included
the natural size of at least $\sim$ (TeV$)^4$ for the vacuum energy
(as well as other sources of energy associated with,
for instance, radiation, matter or the inflaton field).
There are also other terms in this definition of
$T_{\mu \nu}$, for example purely gravitational terms
arising from SM loops with external graviton lines. However, all of these
effects are absorbed into higher-dimension operators in $\Gamma_{\rm SM}$
suppressed by powers
of $M_{\rm Pl}$. Given the enormity of $\bar{M}$ and the consequent
miniscule size of $\bar{R}$, these operators have negligible
effects suppressed by powers of $({\bar R}/M^2_{\rm Pl})$.
The same  stability arguments  apply to the finite-$L$
equation (\ref {1}).

We can also see that our modification of gravity has no other
effect in the $L\to \infty$ limit than to generate a
miniscule {\it apparent} vacuum energy.
We simply take the
result
for $\bar{R}$ and put it back into the equation; we have
\begin{equation}
M_{\rm Pl}^2 \,G_{\mu\nu}\,
%\left ( R_{\mu \nu} - \frac{1}{2} g_{\mu \nu} R \right )\,
= \,\frac{M_{\rm Pl}^2}{M_{\rm Pl}^2 + {\bar M}^2}\,
{\cal E} \,g_{\mu \nu} \,+ \,T_{\mu \nu}^{\rm other}\,.
\label{local1}
\end{equation}
Therefore all other (good) properties of Einstein gravity are completely
maintained. In particular, the standard slow-roll
inflationary cosmology is untouched.
We can have an inflaton with a potential satisfying the usual slow-roll
condition, where the minimum of the
inflaton potential is  ${\cal E}$. All of inflation and the
subsequent evolution of the Universe through reheating,
nucleosynthesis, matter
domination and structure formation would go through
unchanged, and we would match it to a universe that would eventually be
de Sitter with tiny curvature $\sim {\cal E}/\bar{M}^2$.

We can see this in a closely related way again by
tracing and taking the space-time
average of the l.h.s. and r.h.s. of Eq. (\ref {rbar}), we arrive at

\beq
\mpl^2\,G_{\mu\nu}\,=\,T_{\mu\nu}\,-
\,{1\over 4}\,g_{\mu\nu}\,  {{\bar M}^2 \over \mpl^2+{\bar M}^2 }
{\bar T}\,,
\label{Tbar}
\eeq
where ${\bar T}\equiv \int d^4x\sqrt{g}T/\int d^4x\sqrt{g}$.
Here we are subtracting the space-time average of
$T$ on the r.h.s. of the Einstein equation, which has the effect of
subtracting out the vacuum energy in an asymptotically dS universe.
Note that for $\bar{M}^2 \gg
M_{\rm Pl}^2$, the coefficient of
the second term is extremely close to $1$. Had we
simply written down this equation of motion to
begin with, with
$\bar{M}^2/(\bar{M}^2 + M_{\rm Pl}^2)$
replaced by a parameter $x$, then in order
to address the CCP we would have to adjust $x$ to be very close to 1 with
enormous accuracy. (The case with $x=1$ is reminiscent of the proposal of
\cite{tseytlin}). However, what we have seen is that this can be a
consequence of Eq. (\ref{rbar}), where there are no fine adjustments at
all but simply large hierarchies between $M_{\rm Pl}$,
$\bar{M}$ and ${\cal E}$. Since the form of the
equation of motion given in (\ref {rbar}) is free from fine
adjustments, then this is what we should try and match
from a more fundamental theory.

Note that it is impossible to find solutions which are asymptotically
${\it flat}$ in the future; any such solution would have $\bar{R} = 0$,
and be in contradiction with the r.h.s. of Eq. (\ref {local1})
for non-zero ${\cal E}$. Thus,
asymptotically flat spaces can only arise in theories where ${\cal E}=0$,
such as in large classes of supersymmetric models: in those cases of
course no asymptotically de Sitter solutions would exist.

It is interesting to compare our approach  with
``unimodular gravity''. In order to do so, we rewrite Eq.
(\ref {rbar}) as a system of the following two equations:
\beq
\mpl^2\,\left (R_{\mu\nu} \,- \,{1\over 4}\,g_{\mu\nu}\,R
\right )\,=\,T_{\mu\nu}\, - \,{1\over 4}\,g_{\mu\nu}\,T\,,
\label{Eq1}
\eeq
\beq
\mpl^2\,R\,+\,T \,=\,-\,{\bar M}^2\,{\bar R}\,,
\label{Eq2}
\eeq
where $T\equiv T^{\alpha}_{\alpha}$ denotes the trace of the
energy-momentum tensor.

The first equation is that  of unimodular gravity
(see, e.g.,  \cite {Weinberg}).
In addition we get the second
equation, which  plays a vital role. To see this let us
start with  $T_{\mu\nu}\, = \, {\cal E} \,g_{\mu\nu}$.
For this stress tensor the r.h.s. as well as the l.h.s.
of Eq. (\ref {Eq1}) is zero identically.
Therefore, Eq. (\ref {Eq1}) alone does not determine
the curvature. To find the curvature we turn to Eq.
(\ref {Eq2}).  The latter gives Eq. (\ref {cc0}).
Since ${\bar R}$ is a space-time constant
the Bianchi identities are trivially satisfied
for   Eq. (\ref {rbar}), and for the system (\ref {Eq1})--(\ref {Eq2}) too.
We act on both sides of Eq. (\ref {Eq1}) by the covariant derivative
$\nabla^\mu$.  Since the energy-momentum tensor of matter
and the Einstein tensor are covariantly conserved,
$\nabla^\mu T_{\mu\nu} =\nabla^\mu G_{\mu\nu}=0$,
we obtain
\beq
\partial_{\mu} \, \left (\mpl^2\,R\,+\,T \right )\,=\,0\,.
\label{Biancci}
\eeq
This equation implies that $\mpl^2\,R+T$ can be an arbitrary
space-time constant as in unimodular gravity \cite {Weinberg}.
However, the vital ingredient of the present
approach is the second equation,  (\ref {Eq2}),
which uniquely determines the value of
curvature to be in agreement with (\ref {cc0}) in
asymptotically de Sitter space-times.

It is likely that
there are other solutions to our equations of motion, which are not
asymptotically dS; and which for instance correspond to finite-volume
``bang-crunch'' type cosmologies. However, it is clear that
these solutions are not in any sense continuously connected with the
desirable ones that {\it are} asymptotically de Sitter.
Further, as we will discuss below,
there may be more fundamental reasons for restricting ourselves to
asymptotically de Sitter universes.

\subsection{Acausality Instead of Fine-Tuning}

Perhaps the most disturbing feature of our modification to Einstein's
equations is that it is manifestly acausal.
Therefore, it is crucial to understand whether
this may affect observations. We will argue below that the
acausality has no significant effect on
any observable source in the Universe, while
it is very important for solving the CCP.

A simple argument suggests that if some sort of non-local
modification of gravity is responsible for resolving the CCP
in a realistic setting, it should have
acausality as a fundamental feature.
Imagine a {\it causal} modification of
gravity in the infrared, at some
scale $\sim L$, such that while
energy-momentum localised in space or time
over scales smaller than $L$  gravitate normally,
energy-momentum spread
out over scales much larger than $L$ hardly gravitate. For our Universe,
we would have to require that $L$ be at least of the order the size of our
Universe today, $\sim 10^{28}$ cm.
Now suppose
that the Universe begins in a big bang at some time moment
$t=0$; and the energy-momentum tensor has a vacuum
energy component that we are trying to
render
harmless by our IR modification. But how can {\it causal} physics know
whether the cosmological constant is truly {\it constant} and not some
temporary blip, as in inflation? If it {\it were} a blip, which
disappeared
after a time much smaller than $L$, it would have to gravitate
normally, and therefore inflate. It would then take causal physics a time
of order $L$, which is at least ten-billion years,
to recognise that the cosmological constant was truly
constant; only then would the large rate of inflation cease.
But for our Universe, there are observational reasons to believe that this
can not happen. The successes of standard cosmology starting from
big-bang nucleosynthesis at 1 second, to the present, indicate that the
Universe must have been ``normal'' since it was one second old.
There may be loopholes around this
argument, however the issue of the time scale needed to cancel the CC must
clearly be addressed in any scenario with causal modifications of gravity.

By contrast, our non-local modification is maximally acausal in the
interesting case of an asymptotically de Sitter universe, because it is
dominated by the deep {\it future} behaviour of the geometry. However,
this acausality does not lead to any peculiar behaviour (other than the
suppression of the effective CC!), precisely because the
deep future behaviour in a de Sitter space is so universal. In fact, we
have seen that after self-consistently solving Eq. (\ref {rbar})
for $\bar{R}$ and inserting the solution back into the
equation of motion, we have a completely
{\it local} equation of motion (\ref {local1}),
where, however, the vacuum energy part of the stress tensor seems to be
unnaturally small!
Therefore, at least for asymptotically
de Sitter spaces, there are two descriptions of the physics: one which is
free of any fine-tuning but highly acausal, and another
which is {\it local} and {\it causal}, but which appears to have a
highly unnatural value for the vacuum energy.

The acausality also takes care of one of the usual
conundrums associated with
attempting to make vacuum energy ``not gravitate''. Locally in time, we
just have some energy momentum tensor $T_{\mu \nu}$ composed of, say,
contributions from matter and radiation as well as ${\cal E} g_{\mu
\nu}$. How can we
disentangle the ``vacuum energy'' part of
$T_{\mu \nu}$ and tell it not
to gravitate? Local physics can not do this without modifying the
response of gravity to either matter or radiation as well; however, our
acausal modification knows how to do this. The vacuum energy
part of $T_{\mu \nu}$ is precisely
the part that does not dilute away in an
asymptotically de Sitter universe deep in the future.

Finally, the acausality also resolves another minor puzzle: as we have
seen, the $L\to \infty$ of Eq. (\ref {1}) does not
lead to Einstein's theory. This seems to violate
naive decoupling intuition. The physical reason that this does not
violate decoupling is that $L$ is not the true infrared scale
of the theory. There is an even longer time scale,
the (infinite) total  age and size of the asymptotically de Sitter
universe.

\subsection{Possible Connection with dS/CFT}

There is a quite different way of motivating our equation of motion
(\ref {rbar}), which
resonates with some interesting qualitative ideas springing from the
dS/CFT conjecture \cite{dscft}.

Our present formulation of the equations of motion (\ref {rbar}),
involving the space-time average of $R$, gives rise to perfectly
satisfactory physics for asymptotically de Sitter spaces, though, as
we mentioned, likely  there are other solutions that are not
asymptotically dS. Actually, since our
equation is non-local in time in any case, we could declare that we are
{\it only} interested in universes that are asymptotically dS in the
future. This may seem like a perverse thing to do at first, but it is
quite
sensible when viewed in the context of the dS/CFT conjecture. In this
correspondence, there is a dual description of physics in asymptotically
de Sitter spaces, in terms of some sort of Euclidean conformal field
theory
in one lower dimension without gravity. A concrete example of such a
theory
is still lacking, but if one existed, we would be tempted to say that it
is
the fundamental definition of quantum gravity in asymptotically
de Sitter space-times, much as ${\cal N}=4$ SYM is taken to be the
fundamental description of backgrounds in string theory that
are asymptotic  to $AdS_5 \times S_5$.

In this correspondence, time is the holographically generated dimension
and
is associated with the RG scale in the dual theory; with the {\it
ultraviolet}
of the dual field theory identified with the {\it deep future} in the
space-time description. the RG flow from the
UV to IR is associated with inverse time evolution in the bulk.
For instance,
an RG flow in the field theory that starts near a UV fixed point, and
skims by an IR fixed point before developing a mass gap and
becoming free in the deep IR,
plausibly has a space-time description in terms
of a universe that begins
in a big bang, goes through an inflationary epoch (near the IR fixed
point) and ends in an
accelerating dS phase (UV fixed point).
It is fascinating that in this picture, the
fundamental degrees of freedom, corresponding to the deep UV of
the dual theory,  are associated not with
the big bang but with the {\it deep future}
dS phase of the universe.

If we view the history of our own Universe in terms of RG flow in some
dual CFT,
there is then simply no choice about the deep future behaviour of the
Universe; the fundamental (UV) theory is dual to a space-time that is
asymptotically de Sitter in the future. However, the CCP still
exists: Why is the asymptotic de Sitter radius so much larger than the
naive
expectation from the  Standard Model vacuum energy?

As we have seen, we can rephrase the question. The conundrum only
results from assuming that the effective space-time description of this RG
flow is given by the
(local) Einstein equations. But this has not been derived at the
current level of understanding of quantum gravity in de Sitter space in
general, or of dS/CFT in particular.
We can therefore entertain the possibility that
the effective classical space-time description of the RG flow differs from
Einstein's equations, in such a way that the largeness of
the asymptotic de Sitter radius does not appear finely tuned. An obvious
possibility is
\begin{equation}
M_{\rm Pl}^2\, G_{\mu \nu}\, - {1\over 4}\, \bar{M}^2 \,g_{\mu \nu}\,
R_{\infty}\, = \,T_{\mu \nu}\,,
\label{Rinfty}
\end{equation}
where $R_{\infty}$ is now simply defined as the asymptotic dS curvature,
rather than in terms of any space-time average. In fact, $R_{\infty}$ is a
natural quantity in dS/CFT, because the UV
value $c_{\rm UV}$ of the CFT
central charge---essentially the number of degrees of freedom in the
CFT---is determined by the curvature in Planck units as
\begin{equation}
c_{\rm UV} \, = \, \frac{M_{\rm Pl}^2}{|R_{\infty}|}\,.
\end{equation}
At any rate, all the consequences of Eq.  (\ref {Rinfty})
are identical to what we
have already seen, and again, we can deduce an
equivalent description which is local but where the CC appears absurdly
finely tuned. However, the fundamental starting point and interpretation
here are rather different.

Since the dual theory seems to have only one relevant large dimensionless
number, the central charge $c_{\rm UV}$, it is tempting to imagine that the
parameters $\bar{M}$ and the vacuum energy ${\cal E}$ are both determined
in terms of $c_{\rm UV}$, in such a way that the produced
$\bar{M}$  is huge with respect to
the Planck scale, while also producing a small Standard Model vacuum
energy (which  in a supersymmetric theory would be related to the electroweak
scale in the usual way). Such a correlation offers a possible simultaneous
solution to the CCP as well as the hierarchy problem, and may be related to
the proposal of Banks \cite{banks}.
It can also
explain the oft-noted striking ``co-incidence'' that the weak, Planck and
Hubble scales appear to be related as $H^2 \sim m_{\rm EW}^8/M_{\rm Pl}^6 \sim
{\cal E}^2/M_{\rm Pl}^6$. This can arise if
we take $\bar{M}$ and ${\cal E}$ to scale with $c_{\rm UV}$ as
\begin{equation}
\bar{M} \sim c^{1/2}_{\rm UV} \, M_{\rm Pl}, \, \, {\cal E} \sim  c_{\rm
UV}^ {-1/2} \, M_{\rm Pl}^4\,,
\end{equation}
which yields
\begin{equation}
\frac{H^2}{M_{\rm Pl}^2}\, = \,\frac{R_{\infty}}{M_{\rm Pl}^2}\, = \,
\frac{1}{c_{\rm UV}}\, = \,\frac{{\cal
E}^2}{M_{\rm Pl}^8}\,.
\end{equation}

\section{Finite $L$ Theories}

We have seen that addressing the CCP in our approach can already be done
in
the $L \to \infty$ limit, where the only remnant of our
non-local modification, at least in asymptotically de Sitter space-times,
is the suppression of the cosmological constant. But there are clearly
potentially new phenomena associated with finite $L$ theories, and in any
case finite $L$ acts as a regulator that can shed some more light on
the general mechanism for suppressing the CC.

\subsection{Toy Scalar Example}

Let us begin by considering a toy scalar example of a finite $L$ theory,
where the complications of non-linearity and tensor structure of a full
gravitational theory are removed, but where much of the physics
associated with
our non-local modifications remain. This example is essentially the
linearised gravity approximation without the indices. To begin with,
consider a scalar
field $\phi$ coupled to a source $T$. The equation of motion is
\beq
\nabla^2 \phi \,+ \,T \,= \,0\,.
\eeq
We can think of
\beq
R_{\rm toy} \,\equiv \, \nabla^2 \phi\,,
\eeq
as our toy analog of the gravitational
curvature.

In this toy world, the field $\phi$ couples to matter and
mediates  long-range
scalar gravity, which has been successfully ``measured'' to distance
scales $\sim H^{-1}_0$; also the ``gravitation'' of spatially
homogeneous sources
has been measured for times as long as $\sim H_0^{-1}$.
The toy analogue of the cosmological constant problem is that the source
$T$
is also expected to contain a large space-time homogeneous component
${\cal
E}_{\rm toy}$, but the measured value of $R_{\rm toy}$
is far smaller than $-{\cal E}_{\rm toy}$.
This can be remedied by modifying the
equation of motion as
\beq
\left ( 1 + \fu(L^2 \nabla^2) \right )\, R_{\rm toy}\, +\, T\, = \, 0 \,.
\eeq
We would like to ensure that this modification removes the toy CC problem
without adversely affecting anything else. This will require constraints on the
function $\fu$ in addition to the basic requirements that $\fu(\alpha)
\to
0$ for $\alpha \gg 1, \fu(0) \gg 1$.
Passing to  momentum space, we find that
\beq
\left (1 + \fu(L^2 p^2) \right )\, \tilde{R}_{\rm toy}(p)\, + \,
\tilde{T}(p)\, = \, 0\,.
\eeq
At generic momenta we can simply divide by $(1 + \fu(L^2 p^2))$ and
conclude
that
\beq
\tilde{R}_{\rm toy}\, =\, -\frac{\tilde{T}(p)}{1 + \fu(L^2 p^2)}\,.
\eeq
For a constant source, $\tilde{T}(p) = {\cal E} \delta(p)$, and in the
limit where
$\fu(0) \to \infty$, we can make $R_{\rm toy}$ vanish. Phrased
diagrammatically,
we have modified the $\phi$ propagator to vanish at zero external
momentum,
and therefore the $T$ tadpole does not force any space-time variation for
$\phi$, and so $\phi$ can be put at any constant value
$\phi_0$. Furthermore, as long as $(1 + \fu(L^2 p^2))$ rapidly
approaches 1
for real $|p^2|>H_0^2$, we are guaranteed that the
successes of our toy gravity are maintained.

In general, the function $(1 + \fu(z))$ may have zeros in the complex $z$
at some points $z_i$; if so, there will be new solutions to the
homogeneous equations of motion with effective frequency $\omega_i^2 =
\frac{1}{L^2} z_i$.  The existence of such ``transients'' is disastrous
because it implies that there are solutions
\begin{equation}
R_{\rm toy}\, \sim  \, R_0\, e^{-i \omega_i t}\,,
\end{equation}
where $R_0$ is {\it any} initial amplitude. Thus, while it may be
impossible to find solutions for $R_{\rm toy}$ that are
large and {\it exactly}
stationary, there are infinitely many solutions where $R_{\rm toy}$ has
arbitrarily large amplitude, varying very slowly over time scales of order
$\sim L$. In a realistic gravitational setting, these would correspond to
universes that  are, for example, inflating,
but with the rate of inflation very
slowly changing
\footnote{We thank Maxim Perelstein for discussions on this point.}.
To see this, consider a space-time where
$R \sim {\bar R}+H^2 e^{-i \omega t}$;
Here, the ${\bar R}$ is a particular solution
of Eq. (\ref {1}) with a constant energy density  on its r.h.s.,
while  the parameter $H$ can take any value.
Since the ${\bar R}$ is tiny (because ${\bar M}$ is large),
and if  $\omega$
is very small compared to $H$, then in computing
$(1 + \fu(L^2 \nabla^2)) R$, to
a good approximation we
can use the metric with constant $H$ in $\nabla^2$.
Thus,  $\nabla^2 \simeq \partial_t^2 + 3 H \partial_t$
and we obtain
\begin{equation}
(1 + \fu(L^2 \nabla^2))\, H^2 e^{-i \omega t} \,\sim \,
(1 + \fu(-L^2 \omega^2 + 3 i H
L^2 \omega))\, H^2 \,e^{-i \omega t}\,.
\end{equation}
Therefore, if $(1 + \fu(z))$ has roots in the complex plane, there will
be solutions which are inflating with a rate $H$ which is very slowly
changing on a time-scale of $H L^2$ or $L$, whichever is larger.
We would then have to
explain why the Universe is not in one of these solutions,
converting the fine-tuning problem for the CC into an
essentially equally finely-tuned
question about initial conditions. It is therefore desirable to eliminate
such transients by insisting that $(1 + \fu(z))$ have no zeros in the
complex plane (except at infinity).

An explicit example of a function that
has these properties is
\beq
1\,+\,\fu (L^2 p^2) \,=\,{1\over 1 + \epsilon \,-\,e^{-(L^2\,p^2)^2}}\,,
\label{Fex2}
\eeq
which quickly asymptotes to 1 as $|p^2| L^2$ gets large,
satisfies $\fu(0) \to \epsilon^{-1} \simeq {\bar M}^2/\mpl^2 $,
and has no zeros at any finite point in the complex plane.
Another example that satisfies all the above requirements is
\beq
1\,+\,\fu (L^2 p^2)\,= \,{\rm exp} \left ( N e^{-(L^2\,p^2)^2} \right )\,,
\label{Fex3}
\eeq
where $N\simeq {\rm ln}({\bar M}^2/\mpl^2)$ is some large number.

Let us restrict ourselves to sources that are spatially homogeneous.
Then, we can solve for the toy curvature as a function of time
\beq
R_{\rm toy}(t)\, =\, -T_{\rm eff}(t)\,,
\eeq
where
\beq
T_{\rm eff}(t) \,= \,T(t)\, - \,
\int d t^\prime K(t - t^\prime)\, T(t^\prime)\,,
\eeq
with the kernel $K$ given by
\begin{equation}
K(t)\, = \,\int d \omega \,f(L^2 \omega^2)\, e^{- i \omega t} \,.
\end{equation}
Here the function $f$ is defined as $(1 + \fu)^{-1} \equiv (1 - f)$.
Note that since we have assumed that $1 + \fu$ has no zeros, $(1 - f)$ has
no poles on the real axis. Furthermore, $f(\alpha)$
quickly asymptotes to zero for large $\alpha$, while it approaches
$1$ near $\alpha = 0$. Therefore, the integrand in the
definition of
$K(t)$ is regular. Since $f$ is a function of $\omega^2$, $K$ is manifestly
time-symmetric, $K(t) = K(-t)$ and is therefore acausal. Therefore we
conclude that if we wish to
avoid the existence of the undesirable transient solutions, we are {\it
forced} to have acausality. As we have argued in
the previous section, however,
such acausality is actually welcome for a realistic
solution of the CCP.

Note as well, that as $L \to \infty$, $\int
dt^\prime K(t - t') T(t')$ is just $f(0)$ times
the time average of the source $T$. For $f(0)=1$ and a constant $T$,
$T_{\rm eff}$ vanishes. Now consider a source which is a step
function, $T(t) = T_1$ for $t<0$, $T(t) = T_2$ for $t>0$. This is the toy
analog of an inflationary phase transition from
one vacuum energy to another. Here, for
$L \to \infty$ and $ f(0) =1$, we have $T_{\rm eff} = T(t) - (T_1 + T_2)/2$.
The effective source is not cancelled for either $t>0$ or $t<0$. This is
because the source is uniform in both the deep past and the deep future,
and the space-time average is thus equally weighted by  past and future
values, and does not cancel either of them.
This behavior would be a disaster for our real gravitational case of
interest, but
fortunately, our toy example fails as a good analogy here
for the following two reasons.
First, any realistic cosmology has an origin of
time in the big-bang, and therefore
the inflationary potential does not dominate the energy
for infinite times into the past,
while in an asymptotic de Sitter universe the vacuum energy persists
infinitely into the future. The average is then dominated by the deep
future value of the vacuum energy, which is what is effectively cancelled
(or suppressed for large but not infinite $\bar{M}$).
Second, the curved space d'Alambertian, $\bo$, introduces
a new scale in the complete theory---the Hubble parameter of the
corresponding time-dependent background $H$,
or square root of the scalar curvature for
static backgrounds. This scale
enters the argument of the $\fu$ and
plays a vital role in calculations.
Due to the presence of that scale
there is no delay by $L$ in the response of the $\fu$ to any
sudden change in a source, such as, e.g.,
the change during phase transitions.

We now want to consider finite $L$ modifications for real gravity, by
returning to Eq. (\ref {1}).

\subsection{Bianchi Identities}

Let us act on both sides of Eq. (\ref {1}) by $\nabla^\mu$.  Since
$ \nabla^\mu G_{\mu\nu}\,=\, \nabla^\mu T_{\mu\nu}=0$ we get
\beq
\nabla^\mu \left ( \fu (\bo)\,G_{\mu\nu}\right )\,=\,0\,.
\label{bianchi}
\eeq
The covariant derivative does not commute with $\bo$
for general backgrounds; therefore,
(\ref {bianchi}) is an {\it additional}
constraint  on $G_{\mu\nu}$.
Hence, the Bianchi identities are not just kinematically
satisfied as in the Einstein gravity.
Instead, all possible consistent solutions of
Eq. (\ref {1}) should satisfy the new constraint
(\ref {bianchi}). It is of course possible that there is a more clever
non-local modification that kinematically satisfies the Bianchi
identities, but let us press on.
Note that the same conclusion can be derived by
inverting the $1+\fu$
operator in Eq. (\ref {1}) and putting it to the r.h.s. Then,
the Bianchi identities lead to the following constraint
\beq
\nabla^\mu \left ( {1\over 1\,+\,\fu(L^2\bo)}\,T_{\mu\nu} \right )
\,=\,0\,.
\label{bianchi1}
\eeq
For a given source $T_{\mu\nu}$  this equation should be interpreted
as a new constraint on the corresponding metric that enters the covariant
derivatives in (\ref {bianchi1}), and not as
a constraint on the source itself.

Note that the tensor $\tau_{\mu\nu}\equiv
(1\,+\,\fu(L^2\bo))^{-1}T_{\mu\nu}$
is covariantly conserved according to Eq. (\ref {bianchi1}).
Order by order in perturbations,  we  can try to restore
an action that would give rise to $\tau_{\mu\nu}$
as its stress-tensor. There are two important things about that action.
First, it contains an infinite number of local terms that cannot be
truncated at any finite order; therefore,  the would-be action is
fundamentally non-local. Second, we notice that this action contains
additional vertices of interactions of gravitational perturbations with
matter fields in $T_{\mu\nu}$. These vertices may play a crucial role
in certain processes discussed below.

Let us start with  perturbations on a  flat background for which
$g_{\mu\nu}\,=\,\eta_{\mu\nu}\,+\,h_{\mu\nu}$.
We would like to check what are the new constraints imposed
by (\ref {bianchi}) on these perturbations.
In the linearised approximation
Eq. (\ref {bianchi})  is  satisfied if $\partial^{\mu} h_{\mu\nu}=
\partial_\nu h^{\alpha}_{\alpha}/2$.
The latter expression  is nothing but the {\it harmonic} gauge-fixing
condition in
the Einstein gravity. Hence, in
the linearised approximation  the constraint
(\ref {bianchi}) does not give rise to any new restrictions
on perturbations.

As a next step,  we write down the
expression for the response of gravitational field to a source
with the stress-tensor $T_{\mu\nu}$. From (\ref {1})
we derive
\beq
h_{\mu\nu}\,=\,{8\,\pi \,G_N
\over [1+\fu(\bo)]\,\bo }\, \left (T_{\mu\nu}\,-\,{1\over 2}\,
\eta_{\mu\nu}\,T^{\alpha}_{\alpha} \right )\,.
\label{respT}
\eeq
The novelty is the appearance of $(1+\fu)^{-1}$ on the r.h.s.
This  suppresses the gravitational response to
states with $p^2=\omega^2 -{\vec p}^2\,\sim \,0$.
The role of such states in ordinary astrophysical objects,
such as  stars and planets, is negligible.
Therefore, we do not expect any substantial modification
of gravitational fields of classical astrophysical objects.
Note that a massless particle that  is exactly on shell
will generate a
suppressed gravitational field. However, any massless particle that is
emitted and absorbed over scales smaller than $L$ is always off-shell by an
amount greater than $1/L$, and will therefore gravitate normally.

Note also that, in this model, action and reaction are not necessarily
equal. To see this let us calculate the
interaction between sources $T_{\mu\nu}$ and $T^{(1)}_{\mu\nu}$.
In the lowest approximation in $G_N$ the source $T_{\mu\nu}$
sets the gravitational field given in Eq. (\ref {respT}).
The interaction with $T^{(1)}_{\mu\nu}$ is proportional
to $ h^{\mu\nu}T^{(1)}_{\mu\nu}$.
However, the latter expression is not symmetric w.r.t. the
interchange of the sources. Hence, the action of  $T_{\mu\nu}$
does not equal the  reaction of $T^{(1)}_{\mu\nu}$ and
{\it  vice versa}. As in electrodynamics, this indicates that there should
be some radiation that accounts for the mismatch between the action
and reaction. Let us call it the {\it L-radiation}.
Perhaps this is gravitational radiation
due to the new vertices that appear in the covariantly conserved
stress tensor $\tau_{\mu\nu}\equiv (1\,+\,\fu(L^2\bo))^
{-1}T_{\mu\nu}$, which we discussed above.
Alternatively, this may be interpreted as a new form of radiation
of states that should be ``integrated in'' in order to
make our action local.  The vacuum energy
$\cal E$ in our case does not act on  matter in the
Universe since the former sets no
gravitational field.  However, the matter does act on the
vacuum. As a result, there should be the compensating
$L$-radiation from the vacuum, especially  in the regions
in the Universe  that have  high matter density.

Finally all known non-linear solutions should satisfy (\ref {bianchi}).
For the Schwarzschild solution,  $G_{\mu\nu} =0$ everywhere outside
the source. Therefore it trivially satisfies (\ref {bianchi}).
Inside the source, however, the standard solution will be modified
by a quantity that vanishes in the limit $L \to \infty$.
Equation (\ref {1}) has also an exact dS-Schwarzschild
solution outside of the source \footnote{We thank Nemanja
Kaloper for pointing this out to us.}.
The gravitational radius in that solution
is proportional to the conventional Newton coupling $G_N$,
while the curvature is proportional to an effective coupling
$G_N/(1+\fu(0))$.

\subsection{Standard Astrophysics and Cosmology}

Since our proposal modifies gravity, it is important to ensure that it
does not alter standard astrophysics and cosmology. As long as the scale
$L$ is larger than the present size of the observable
Universe $L\gsim 10^{28}~{\rm cm}$, it is clear that
astrophysics will not be affected because the
relevant length scales are shorter than $L$ and consequently the filter
function vanishes. Similarly, early cosmology will not change
because the relevant length
scales---horizon size, particles' Compton wavelengths,
and inverse temperature---are all shorter than $L$.

Mathematically we can see this by decomposing the stress tensor into
its vacuum energy piece and the rest (matter plus radiation):

\beq
T_{\mu\nu}\,=\,{\cal E}\,g_{\mu\nu}\,+\,{T}^{\prime}
_{\mu\nu}\,,
\label{Ttilde}
\eeq
where ${T}^{\prime}_{\mu\nu} $ denotes the stress tensor
for everything but the vacuum energy.
The solution of Eq. (\ref {1})
with (\ref {Ttilde}) on the r.h.s. takes the form
\beq
G_{\mu\nu}\,=\,\Lambda\,g_{\mu\nu}\,+\,{G}^{\prime}_{\mu\nu}\,,
\label{Gtilde}
\eeq
where $\Lambda$ is defined by
\beq
\Lambda \,\equiv \,
{{\cal E}\over \mpl^2\,[1+\fu(0)]}\,\ll {{\cal E}\over \mpl^2\,}\,,
\label{Lambda}
\eeq
and ${G}^{\prime}_{\mu\nu} $ satisfies the equation
\beq
\mpl^2\,{G}^{\prime}_{\mu\nu}\, \simeq \,{T}^{\prime}_{\mu\nu}\,.
\label{GTtilde}
\eeq
To obtain the latter expression
we used  $\fu(L^2\,\bo) {G}^{\prime}_{\mu\nu} \simeq
\fu({\rm argument}\gg 1) {G}^{\prime}_{\mu\nu}\simeq 0$.
Equation (\ref {GTtilde}) is the conventional Einstein equation on
the background with the vanishingly small
cosmological constant (\ref {Lambda}).

So, the pure vacuum energy gravitates with
$G_N^{\rm eff}\equiv G_N/(1+\fu(0))\ll G_N$,
while the matter and radiation  in the Universe
gravitate with the conventional Newton constant $G_N$. As a result,
early cosmology remains unaffected, while eventually the small gravity
of vacuum takes over and dominates the dynamics of the Universe.  Since
we do not have a theory predicting the value of the filter function,
we cannot explain the observational fact why this is happening in our
epoch.

\subsection{Inflation and Exit}

It is natural to worry that theories addressing the CCP
might have unwanted consequences for inflation
\cite {Guth,Linde,Steinhardt,LindeChaotic,Lindebook}.
First, inflation is driven by vacuum energy which may not
gravitate in such theories.
Second, inflation ends when the Universe transitions
to the ``true vacuum'' which is normally
assumed to have zero energy --- an assumption
that is now replaced by a dynamical principle.
In this section we argue that our proposed
modifications of general relativity maintain
the successes of some of the existing inflationary
scenaria --- notably ``new'' and chaotic inflation---
while suggesting new possibilities for simple and perhaps
more natural inflationary theories.

It is clear that our modified equations differ from Einstein's only
for systems that are simultaneously slow and big compared to $L$.
Conversely, systems that are either fast or small compared to $L$
 behave according to the familiar laws of general relativity.
An example of such a system is the Universe inflating according
(now old) ``new inflation'' paradigm \cite{Linde,Steinhardt}.
There, the inflationary phase is driven by a scalar field
rolling with a characteristic time scale much shorter than $L$,
and is therefore unaffected by our modifications. The same is true
for chaotic inflation \cite{LindeChaotic}  in which an overdamped
scalar field rolls, again at a rate fast compared to
$L^{-1}$. The subsequent phases of exiting inflation and
reheating also occur on time-scales shorter than $L$.
So, the only effect of our modification, in either
new or chaotic inflation,  is to ``filter out''
the (arbitrary) constant part of the potential, as desired.

To find a novel inflationary scenario --- one where our
modifications make a difference--- we first have to look
for a system that is slow compared to $L$. A simple example is a
classical scalar field at rest at some local minimum of
its potential (not necessarily the true minimum). This
suggests the following possibility: Consider a false vacuum
bubble (or island) of size $d$ and positive energy density $V$,
created at time $t=0$ by tunnelling from the true vacuum.
The Fourier components of such a potential at time $t=0$ have
characteristic wavelengths of order $d$, the initial size of the system.
Suppose that $\mpl/\sqrt{V} \ll d \ll L $.
Then, the region of space within the island
will start to inflate with the conventional rate
and the characteristic wavelengths of the system will get red-shifted,
and---one by one, longer ones first--- will get
stretched beyond $L$. As a result,
these wavelengths become decoupled from gravity or ``degravitated''.
This gradual peeling-off  of the gravitating
Fourier components ends when the whole  potential
has been degravitated causing inflation to terminate.
The above mechanism of ``self-termination'' for exiting inflation
is inherent in our framework. It naturally leads to the
final state of very weakly gravitating vacuum energy.

To decorate these ideas with equations, let us study
the behavior of the scalar curvature during inflation.
The trace of the modified Einstein equation takes the form
\beq
-\mpl^2 \, \left (1\,+\,F(L^2\,\bo) \right ) \,R
(t, {\vec x}) \, = \,T(t, {\vec x})\,.
\label{trace11}
\eeq
Concentrate now on a small region of size
$|\Delta x|\ll d$ inside the island.
Since the boundary effects are negligible in that region,
we expect that the curvature there is approximately
constant (to an accuracy  of ${\cal O} (|\Delta x|/ d)$).
Therefore, the metric in that region
can be approximated  by the standard ansatz,
$ds^2 =dt^2-a^2(t)d{\vec x}^2$.
Then, from (\ref {trace11}) we obtain
\beq
R \,\simeq \,- {4V\over \mpl^2}\,
\left \{ 1\,+\,\fu \left (
{L^2 \over d^2 a^2} \right ) \right  \}^{-1}\,,
\label {RV}
\eeq
where we used the fact that the characteristic (covariant)
momentum square in the bubble of the initial size
$d$ scales as $|k_*^2|\sim g^{ii}/d^2 = 1/d^2a^2(t)$.

As long as the argument of $\fu$
is larger  than unity, we are back to
general relativity.
However, if the scale factor $a$ grows
with time, there comes a moment  $t=t_0$ after which  $ L\lsim d a(t_0)$.
Thus, for $t>t_0$ the argument of the $\fu$
function drops below unity and the denominator of Eq.
(\ref {RV}) becomes enormous. As a result, for $t\gsim t_0$
inflation continues with a very small
rate that is suppressed  by $\fu(0)$.
Since for an inflationary potential $a\sim {\rm exp}(Ht)$,
the period of rapid inflation ends when
\beq
t\,\gsim \,t_0\,\sim \, {1\over H}\, {\rm ln}(k_*\,L)\,,
\label{t0}
\eeq
where $k_*$ denotes the characteristic
physical momentum  in the initial state of the system.
To summarise: the system undergoes rapid inflation for the period of time
$0<t<t_0$,  after which it settles to the reduced inflation rate:
\beq
H_{\rm reduced}\,=\,\left
(V\over \mpl^2\,[1+\fu (0)]\right)^{1/2}\,.
\label{Hred}
\eeq
The present-day acceleration of the Universe determines $\fu(0)$.

So far we have shown how degravitating can lead to a new exit
from inflation that is inherent in our framework and does not
make use of  the detailed shape of the potential.
A realistic inflationary model must also provide
a mechanism for reheating, to produce matter in the Universe.
One possibility that is generic is reheating due to
particle creation \cite{Ford}. If the change in curvature
due to our mechanism is abrupt this results in the production of
pairs of particles. The pairs can reheat the Universe provided that
they are not redshifted by the expansion.
From Eq. (\ref {t0}) we deduce that the time scale
for changing  the  curvature is
\beq
\Delta t \,=\,{1\over H}{|\Delta k |\over k_*}\,,
\label{dedek}
\eeq
where $|\Delta k |\sim  1/d$ is the band of wave-vectors
associated with the  initial island.
To avoid total redshift of the created pairs,
$\Delta t $  has to be shorter then the
doubling time of the universe, $\Delta t < H^{-1}$.
Combining this with (\ref {dedek}) we find the mild condition
$|\Delta k|   < k_*$.

Therefore, in our framework, if this condition is satisfied,
the island (or bubble) of constant potential provides us with a
scenario that has inflation, exit, reheating as well as eternal
acceleration at the reduced rate of Eq. (\ref{Hred}).
Of course, the arguments presented above are a sketch of  real
computations that have to be done properly
to see if these ideas are completely viable.

Note that in the conventional approach  this scenario would have been
impossible since there is no homogeneous classical process in
general relativity
that  would end inflation and lead to reheating in such a uniform island.
One  conceivable mechanism is inhomogeneous and involves
quantum-mechanical bubble nucleation, as was proposed by
Guth in his original inflationary scenario \cite {Guth}.
Unfortunately, within the framework of conventional general relativity
this  scenario is excluded \cite {GuthWeinberg}.
However, we can use the degravitation mechanism described above
to end Guth's inflation without invoking
the bubble nucleation process that causes problems in
Guth's original proposal. If this is the way
things are, we still live inside an island of
false vacuum and all matter in the Universe
originated in the Hawking radiation that got
converted into matter by the sudden
self-termination of inflation.

\newpage

\section{Discussion and Outlook}

It is likely that the ultimate solution to the cosmological constant
problem will require a major shift from some currently cherished physical
principles. Given that the problem is associated with far-infrared
scales, locality is a natural target to be sacrificed. But how can this
concretely address the problem, and why does the world appear at least
approximately local? In this paper we have attempted to address these
questions at a long-wavelength, classical level,
by presenting simple, non-local modifications of Einstein's equation that
dramatically weaken the gravitational effect of vacuum energy, while
preserving the usual successes of general relativity, so that
a natural vacuum energy of size $\sim$ (TeV$)^4$ or even
$(10^{19}~{\rm GeV})^4$ does not lead to
unacceptably large curvature.
The modification can
be qualitatively thought of as making the Planck scale enormous for Fourier
modes with wavelength larger than some scale $L$. This weakens the effect of
sources of energy-momentum uniform in space and time, like the vacuum
energy, while leaving the gravitational effect of other sources
unaffected.

In the $L \to \infty$
limit,  the
Einstein equation is modified in a universal way, by the addition of a term
$\bar{M}^2 g_{\mu \nu} \bar{R}$, where $\bar{R}$ is
the space-time average of the Ricci scalar. This term is not
only non-local but also acausal. Nevertheless, in a broad class of
space-times which become asymptotically de Sitter in the future, the entire
effect of this modification is absorbed into making the asymptotic
de Sitter curvature tiny.
The acausality is a crucial ingredient, because it
is the infinite asymptotic de Sitter future that dominates $\bar{R}$ and
leads to a suppression of the asymptotic dS curvature. However,
while the fundamental equations have a large
non-local piece and a natural size for the Standard Model vacuum energy,
in asymptotically de Sitter space-times there is an equivalent description
of the physics which is local but where the effective vacuum energy
appears un-naturally small.
This example shows explicitly how a non-local effect might address the CCP,
while having the rest of physics look local,
reproducing all the usual successes of general relativity.
Of course, in order for the curvature to be sufficiently small,
the scale $\bar{M}$ must
be extremely large, $\bar{M} \sim 10^{48}$ GeV for (TeV$)^4$ vacuum energy
density, or the mass of
the Universe $\bar{M}  \sim 10^{80}$ GeV, for Planckian vacuum energy
density. However, this large value is
stable under Standard Model radiative corrections, which do not generate
non-local operators with enormous coefficients. In this set-up, it is
possible to imagine a common solution to both the CCP and the hierarchy
problem, if the same physics simultaneously generates the enormous
$\bar{M}$ and the tiny supersymmetry breaking with respect
to the Planck scale.

For  finite $L$, a host of exciting new physical phenomena
arise, including new possibilities for inflation and the exit
from inflation as inflating bubbles stretch to sizes greater than $L$ and
stop gravitating. This opens up new directions to explore both in the
context of standard inflationary scenarios such as new inflation, chaotic
or hybrid inflation, and motivates a re-examination of Guth's old inflation
scenario as well. Further exploration of these ideas, even at this
phenomenological level, could lead to observable consequences that can be
looked for in the CMBR.

There are clearly a large number of avenues to explore. Theoretically, the
obvious outstanding issue is to find a fundamental theory that reproduces
our equation of motion in a classical, long-wavelength approximation. Such
a theory must somehow incorporate the ingredients of non-locality and
acausality in a consistent way, and also lead to an understanding of  the
large size of $\bar{M}$ with respect to
the Planck scale. But even at the level of
our phenomenological equations of motion, many issues need to be
settled. For instance, what do the solutions of Eq. (\ref {rbar})
look like which are not asymptotically de Sitter?
How do we sensibly define $\bar{R}$ in universes with a complicated global
structure? This question is particularly relevant
in inflationary scenaria, once quantum effects for the matter
fields are taken into account, which allow quantum fluctuations of the
inflaton back up its potential hill leading to ``eternal
inflation''. Another question along the same lines is: what happens to
false vacuum inflation once tunnelling is taken into account?
There are similar and even more interesting questions
for the finite-$L$ scenario, together with possibly observable
phenomenological consequences to be explored.

\vspace{0.2in}

{\bf Acknowledgements}
\vspace{0.1cm} \\

We would like to thank Allan Adams, Andy Cohen, Cedric Deffayet, Andrei
Gruzinov, Slava Mukhanov, Lev Okun', Misha Shifman, Eva Silverstein,
Andy Strominger,
Arkady Vainshtein, Alex Vilenkin, Misha Voloshin and Mattias Zaldarriaga
for useful discussions. We would especially like to thank Nemanja Kaloper
and Maxim Perelstein for stimulating discussions on a number of topics.
GG also thanks the Aspen Center for Physics  for hospitality.
GD and NAH are supported in part by a David and Lucille
Packard Foundation Fellowship for  Science and Engineering,
and by Alfred P. Sloan foundation fellowship. GD is also supported by
the NSF grant PHY-0070787. The work of SD is supported by NSF grant
PHY-9870115.

\end{document}